\documentclass[prl,aps,a4paper,twocolumn,showpacs]{revtex4}

\usepackage[ansinew]{inputenc}
\usepackage{array}
\usepackage{color}
\usepackage{amsmath}
\usepackage{amsxtra}
\usepackage{amstext}
\usepackage{latexsym}
\usepackage{dsfont}
\usepackage[pdftex]{graphicx}
\usepackage{epstopdf}
\DeclareGraphicsExtensions{.pdf,.eps,.png,.jpg,.mps}
\usepackage{epsfig}
\newcommand{\ot}{\leftarrow}
\newcommand{\eqr}[1]{Eq.~(\ref{#1})}

\begin{document}

\title{Tunable Supersolids of Rydberg Excitations Described by Quantum Evolutions on Graphs}

\author{P.-L. Giscard$^1$, and D. Jaksch$^{1,2}$}
\affiliation{${}^1$University of Oxford, Department of Physics, Clarendon Laboratory, Oxford OX1 3PU, UK \\
${}^2$Centre for Quantum Technologies, National University of Singapore, 3 Science Drive 2, Singapore 117543}

\date{\today}

\begin{abstract}
We show that transient supersolid quantum states of Rydberg-excitations can be created dynamically from a Mott insulator of ground state atoms in a $2$D optical-lattices by irradiating it with short laser pulses. The structure of these supersolids is tunable via the choice of laser parameters. We calculate first, second and fourth order correlation functions as well as the pressure to characterize the supersolid states. Our study is based on the development of a general theoretical tool for obtaining the dynamics of strongly interacting quantum systems whose initial state is accurately known. We show that this method allows to accurately approximate the evolution of quantum systems analytically with a number of operations growing polynomially.
\end{abstract}

\pacs{67.80.kb, 02.10.Yn}
\maketitle
The quality of control over atomic systems in state of the art experiments is such that one can now address and observe quantum evolutions of individual atoms in optical-lattices \cite{Bakr, Bloch}. Additionally, coherent inter-atomic and light-matter interactions can be made strong enough to occur on  short time-scales compared to incoherent processes. This reveals the system's unitary evolution at the individual constituent level which is of fundamental interest in the study of many-body quantum phenomena. Several applications like quantum simulation and quantum computing schemes also rely on this information \cite{Lewenstein2007}. The relatively new interest in such non-equilibrium dynamics of many-body systems poses serious theoretical difficulties due mainly to the exponentially growing Hilbert space. This in turn hampers advances in the understanding of many-body phenomena such as the elusive supersolids. This new phase of matter was first suggested to exist in Helium as the simultaneous existence of both diagonal and off-diagonal long-range order \cite{SupHelium}. Recent studies have shown that they might be obtainable in optical-lattices \cite{Scarola2006, SupMolecule}.

In this letter we propose the transient creation of supersolid quantum states of Rydberg excited atoms from a Mott insulator in an optical lattice. These are formed by strong laser driving in the presence of long-range dipole-dipole interaction. The interaction causes the excitation probability of an atom to either be inhibited (blockade) or be enhanced (antiblockade) depending on the presence of a nearby Rydberg-excited atom \cite{Jonhson2010}. 
Previously quantum computation and simulation schemes using similar effects have been proposed \cite{Jaksch2000, Ryabtsev2005, Muller2009, Saffman2010, Weimer2010}. Subsequent studies have shown that crystal-like dispositions of the Rydberg excitations could form the ground states of certain 1D and 2D lattices \cite{Weimer2008, Ji2011, Lesanovsky2011} and could also be created dynamically \cite{Weimer2010b, Schachenmayer2010, Pohl2010, Viteau2011}. Yet these studies have not found supersolidity and have been confined to specific parameter regimes. Consequently, the behavior of the system in the general case remains largely unknown and thus we here develop a novel tool for exploring driven evolutions in strongly interacting many-body systems. This is based on summing walks performed by an arbitrarily chosen subset of the system. In the following we call this method walk-sums (WS).

For a system with $N$ constituents each with $d$ internal levels the number of operations involved in computing the evolution-operator $U$ scales as $d^{3N}$. This results from first the number of matrix-elements of $U$ which scales as $d^{2N}$, and second the number of operations required to obtain any one of these with a given accuracy, which scales as $d^N$. With WS it is possible to approximate analytically any chosen pieces of $U$ without any prior knowledge about the system under study and with a polynomial number of operations in $N$. The WS method thus solves the problem of the second exponential scaling while it also bypasses the first one by generating \emph{independently} any desired piece of $U$. For physical quantities requiring an exponentially large number of matrix-elements of $U$ to be calculated, WS can provide approximations by computing randomly chosen pieces of $U$ or only the most relevant ones. Therefore we expect WS to work well for gapped systems where some configurations are unlikely to be populated and can be neglected. WS are also more precise than estimates based on truncations of the Hilbert-space as it can take into account the effect of virtual transitions through configurations outside of the truncated Hilbert-space. Additionally, the number of operations required per element of $U$ remains exponentially better with WS than that of a truncated evolution-operator. Finally, WS provides a reliable way of getting the probability amplitudes of rare events typically inaccessible to Monte-Carlo methods. One can indeed evaluate the dynamics of a specific piece of the wave function and concentrate the computational effort on obtaining a very high accuracy on this single piece. The reasoning we present here is a special case of a very general procedure for working out elements of general matrix functions by summing paths on a graph \cite{Giscard2011b}.


WS are based on splitting a many-body system into a set of constituents $S'$ whose dynamics is frozen and then computing the evolution of the remaining few particles $S$. The surrounding $S'$ being perfectly known, all interactions with $S$ can be exactly evaluated and $S$ evolves through a small Hamiltonian depending on the configuration of $S'$. By expressing the true many-body dynamics in terms of such simple situations with a frozen $S'$ and a few evolving constituents $S$ one can solve the Schr\"{o}dinger equation for large systems. This might seem similar in essence to mean field theory where an atom of interest interacts with a field resulting from the mean behavior of all other particles. The difference is that here we make the mapping from many-body to few-body dynamics exact, that is we make $S$ interact with \emph{all} possible fields it could be subjected to depending on the configuration of $S'$. This will effectively perform some average and mean field behaviors will be recovered but fluctuations around this mean will also be present.

We project the system onto a specific configuration of all constituents in $S'$ by applying the operator $\hat \varepsilon_{\mu}=|\mu\rangle\langle\mu|\otimes \mathcal{I}_{s}$ called a projector-lattice. Here $\mu$ denotes a basis state configuration in $S'$ and the identity is applied to $S$. This operator satisfies the closure relation $\sum_\mu \hat \varepsilon_\mu=\mathcal{I}$ which we insert into the system evolution operator $U$, written as a product of infinitesimally small steps in time $\delta t$. This leads to $U(t)=\sum_\mu \hat \varepsilon_\mu\lim_{\delta t\rightarrow0}\prod_0^m \{\sum_\mu \hat \varepsilon_\mu\} \delta U$,
with $\delta U=U(\delta t)$ and time $t=m \delta t$.
Expanding this product yields terms like
\begin{eqnarray}
\label{eq:Uexpanded}
\hat \varepsilon_{\nu}U(t) &=& \lim_{\delta t\rightarrow0}\hat \varepsilon_{\nu}\delta U\hat\varepsilon_{\nu}\delta U\hdots \hat\varepsilon_{\nu}\delta U + \nonumber\\
&&\lim_{\delta t\rightarrow0}\hat \varepsilon_{\nu}\delta U \hdots \hat \varepsilon_{\nu} \delta U \hat \varepsilon_{\mu}\delta U \hdots \hat \varepsilon_{\mu}\delta U +\hdots
\end{eqnarray}
The first term in this expansion evolves the system from 0 to $\delta t$, then $\hat \varepsilon_{\nu}$ projects $S'$ onto $|\nu\rangle$, followed by evolution for $\delta t$, etc. This freezes $S'$ by continuous (Zeno) measurement of $\hat\varepsilon_{\nu}$ in the limit $\delta t\to 0$, while $S$ evolves freely. The other terms in this expansion describe any number of consecutive Zeno measurements of different configurations of $S'$ switching at all possible times. For instance, the second term of \eqr{eq:Uexpanded} corresponds to one change from configuration $\mu$ to $\nu$. Thus, in this expansion we consider $S$ to evolve in an environment $S'$ which evolves stroboscopically and simultaneously through all possible configurations.

Provided that $S'$ evolves from an initial configuration $\mu$ to a final configuration $\nu$ we thus write the evolution operator for sub-system $S$ as $\hat \varepsilon_\nu U(t) \hat \varepsilon_\mu=U_{\nu\ot \mu}(t)\otimes |\nu\rangle\langle\mu|$ with ($\hbar=1$)
\begin{eqnarray}
\label{eq:Uremarkable}
U_{\nu\ot \mu}(t) =\sum_{n}i^{-n}\sum_{W_n(G)}\int_{0}^{t}\int_{0}^{t_n}\hdots\int_{0}^{t_{2}}e^{-iH_{\nu}(t-t_n)}\nonumber \\
H_{\nu \ot \eta_{n-1}}e^{-iH_{\eta_{n-1}}(t_n-t_{n-1})}\hdots H_{\eta_1\ot \mu} e^{-iH_{\mu} t_1}dt_{1}\hdots dt_{n}.\thinspace
\end{eqnarray}
We note that this expression only contains $d\times d$ matrices. The index $n$ indicates the number of jumps undergone by $S'$ between $0$ and $t$ and the sum over $W_n(G)=\{\eta_0 \equiv \mu, \eta_1 \cdots \eta_{n-1}, \eta_n \equiv \nu\}$ contains all possible strings of $n$ consecutive jumps starting at configuration $\mu$ and ending at $\nu$. The integrals represent continuous sums over all the possible jumping times with one integral per jump. The $d\times d$ matrices $H_{\eta_j}$ are effective Hamiltonians evolving $S$ for a given configuration $\eta_j$ of $S'$ while the matrices $H_{\eta_j\ot\eta_{j-1}}$ describe the effect of a jump on $S$. They are given by
$
\hat \varepsilon_{\eta_j} H\hat \varepsilon_{\eta_j}=|\eta_j\rangle\langle\eta_j| \otimes H_{\eta_j}, ~\hat\varepsilon_{\eta_{j}} H \hat\varepsilon_{\eta_{j-1}}=|\eta_{j}\rangle\langle\eta_{j-1}|\otimes H_{\eta_j\ot\eta_{j-1}}.
$
The operator $U_{\nu \ot \mu}(t)$ is thus the conditional-evolution operator for $S$ \emph{knowing} that initially $S'$ was in state $|\mu\rangle$ and in state $|\nu\rangle$ at time $t$. The time integrals of Eq.(\ref{eq:Uremarkable}) are convolutions and as a consequence the expression of conditional-evolution operators in the Fourier domain only involves additions and multiplications of $d\times d$ matrices
$\tilde{M}_{\eta_j}(\omega)=\mathcal{FT}[\theta(t)e^{-iH_{\eta_j} t}]$,
with $\theta(t)$ the Heaviside function and $\mathcal{FT}$ the Fourier transform. In the Fourier domain Eq.(\ref{eq:Uremarkable}) becomes
\begin{equation}
\label{eq:UFOURIER}
\tilde{U}_{\nu \ot \mu}=\sum_{n}i^{-n}\sum_{W_n(G)}\tilde{M}_\nu H_{\nu \ot \eta_{n-1}}\tilde{M}_{\eta_{n-1}}\hdots H_{\nu_1\ot\mu}\tilde{M}_\mu.
\end{equation}
In the case of $2\times 2$ matrices, we find the $\tilde{M}_{\eta_j}$ to have a simple universal expression in terms of the $H_{\eta_j}$ and that all matrix elements of any product of these $\tilde{M}_{\eta_j}$ are ratios of polynomials with analytically known roots. It follows that we can always analytically perform the inverse Fourier transform back into the time domain for $2\times 2$ matrices. We will see that this formulation of the dynamics is efficient, but a trade-off is that we obtain exponentially little information. Indeed, we really compute $\varepsilon_\nu U(t)\varepsilon_\mu$ i.e.~only $d^2$ out of $d^{2N}$ elements of the full evolution-operator $U$. Computing the conditional-evolution operators for a large number of final configurations is possible because we have analytic expressions, and we can therefore approximate a large number of pieces of $U(t)$. We will follow this approach for Rydberg excitations below.

While every element of \eqr{eq:UFOURIER} can be worked out efficiently it will in general be difficult to distinguish all the possible strings of jumps. We thus proceed by mapping the possible strings of jumps $W_n(G)$ to walks on a graph $G$. We construct a graph $G$ as follows (i) for each configuration $\eta_j$ we draw a vertex $v_{\eta_j}$, and (ii) for each $H_{\eta_j\ot\eta_i} \neq 0$ we draw an edge between vertices $v_{\eta_j}$ and $v_{\eta_i}$. Now all the possible successions of jumps between the initial and final configurations of $S'$ correspond to all the possible walks $W_n(G)$ on $G$ between the initial and final vertices $v_\mu$ and $v_\nu$. The length of a walk is equivalent to the number of jumps in $S'$. We note that it is possible to derive \eqr{eq:Uremarkable} directly from the power-series expansion of the matrix-exponential of the Hamiltonian by exactly summing terms of the form $H_{\eta_j}^n$ which appear as loops on the vertices of $G$ \footnote{Additional cycle resummations further reduce the computational effort \cite{Giscard2011b}.}. The removal of these loops makes our expansion different from a pure power-series expansion. It leads to a significant speed-up when calculating individual elements $\tilde{U}_{\nu \ot \mu}$ and guarantees that a truncation of \eqr{eq:UFOURIER} at order $K$ is at least as accurate as a similar truncation of the power-series. To make this statement quantitative, we compare the number of floating point operations required per matrix element $\varrho$ with that of the power-series expression of $U$. We obtain the number of walks of a given length $K$ between two vertices of a graph from its adjacency matrix $A_G$ \cite{Biggs1993}. The contribution to \eqr{eq:UFOURIER} of order $K$ requires the multiplication of $(2K+1)\langle \nu| A_G^K|\mu\rangle$ matrices with $\langle\nu| A_G^K|\mu\rangle$ the number of walks of length $K$ which is upper bounded by $\sum_z^q \left( N\atop z\right)(d-1)^z$. For most physically relevant Hamiltonians we found $q=1,2$. Therefore $\varrho<d^{-2}\left(2Kd^3+d^2\right) \{q\left(N \atop q\right)(d-1)^{q}\}^K$, where $(2Kd^3+d^2)$ is the number of operations required to multiply the $2K+1$ matrices of a walk of length $K$ and to add the result to the other walks. We compare this to the corresponding value of the Taylor-series expansion $\varrho_T=(K-1)d^{3N}/d^{2N}\simeq Kd^{N}$, and since $\varrho$ is \emph{polynomial in $N$}, the ratio $\varrho/\varrho_T\rightarrow0$ with increasing size $N\rightarrow \infty$.
\begin{center}
\begin{figure}[t!]
\includegraphics[width=.46 \textwidth]{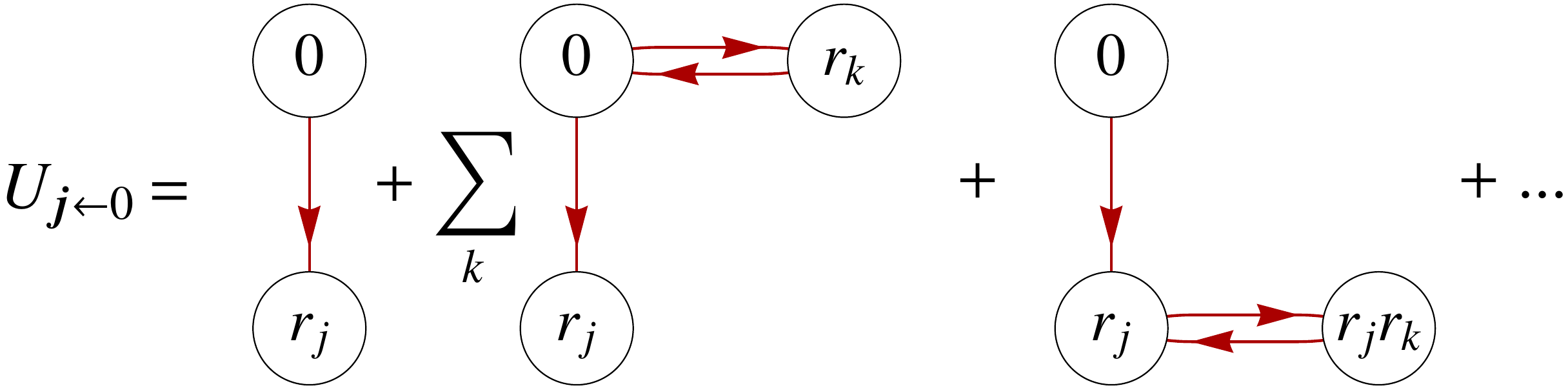}
\caption{Walks of order 1 and 3 contributing to $U_{\bf{j}\ot0}(t)$, the conditional-evolution operator between the initial configuration with no excitations $\ell=0$ (vertex 0) and the final configuration with one excitation $\ell=1$ on atom $j$ (vertex $r_j$). The corresponding mathematical operations are $\tilde{U}_{\bf{j}\ot0}=\tilde{M}_{\bf{j}}H_{1\ot0}\tilde{M}_0+\sum_k\{\tilde{M}_{\bf{j}}H_{1\ot0}\tilde{M}_0H_{0\ot1}\tilde{M}_{\bf{k}}H_{1\ot0}\tilde{M}_0+\tilde{M}_{\bf{j}}H_{1\ot2}\tilde{M}_{\bf{jk}}H_{2\ot1}\tilde{M}_{\bf{j}}H_{1\ot0}\tilde{M}_0\}+\hdots$ }
\vspace{-5mm}
\label{walks}
\end{figure}
\end{center}

We now apply WS to lattices of strongly interacting Rydberg atoms. Initially the atoms are assumed to be in a pure Mott insulating state of the form $|gg\hdots g\rangle$. The atoms are arranged in a regular 2D pattern by trapping them in a deep optical lattice with one atom in internal ground state $|g\rangle$ occupying each lattice site. A laser drives the atoms to highly excited Rydberg states $|r\rangle$ which are strongly interacting over long distances via a dipole-dipole interaction of the form \cite{Jonhson2010, Jaksch2000}
\begin{equation}
A_{ij}=(4\pi\epsilon_0R_{ij}^3)^{-1}[\boldsymbol{\mu_i}.\boldsymbol{\mu_j}-3R_{ij}^{-2}(\boldsymbol{\mu_i}.\boldsymbol{R_{ij}})(\boldsymbol{\mu_j}.\boldsymbol{R_{ij}})],
\end{equation}
with $\boldsymbol{\mu_i}$ the dipole moment of atom $i$ and $\boldsymbol{R_{ij}}$ the relative distance between atoms $i$ and $j$. The laser and dipole-dipole interactions we consider are in the MHz-GHz range and induce dynamics fast compared to incoherent processes and the motion of the atoms. These will limit the lifetime of Rydberg excited quantum states created by fast laser pulses to several $\mu$s but can safely be neglected on the much shorter time scales considered here \cite{Wilk2010}. In this limit the atoms are described by the Hamiltonian
\begin{equation}
\label{eq:TrueHamiltonian}
H=\sum_{i}\{\Delta P_{i} - \frac{\Omega}{2}\left(T_i+T_i^\dagger\right)+\sum_{j\neq i}A_{ij}P_{i}P_{j}\},\vspace{-2mm}
\end{equation}
with $\Delta$ the laser detuning, $\Omega$ the Rabi-frequency, $T_i=|g\rangle_{i}\langle r|$ and $P_i=|r\rangle_i\langle r|$. We choose $S$ to be the atom at the center of the lattice and construct a graph $G$ where each vertex represents a configuration with $\ell$ Rydberg excitations. The Hamiltonian drives transitions $\ell \to \ell \pm 1$ and so $G$ is a linear graph with the vertex representing no excitations at its end. All $H_{\ell\pm 1\ot \ell}=-(\Omega/2) \mathcal{I}_s$ while the Hamiltonian with $\ell$ excitations at positions ${\bf j}=\{j_1 \hdots j_\ell\}$ in $S'$ is given by
\begin{equation}
H_{\bf j}=\mathcal{I}_s(\ell \Delta + \sum_{l,k} A_{j_lj_k})  +\hspace{-.5mm} \begin{pmatrix} 0 &-\Omega/2\\-\Omega/2&\Delta +\sum_l A_{Sj_l}\end{pmatrix}.\vspace{-2mm}
\end{equation}
Figure \ref{walks} shows an example of how we sum walks on $G$ to determine elements of $U$. We calculate all conditional-evolution operators whose final configuration has up to six simultaneous excitations anywhere in the lattice. This gives analytical approximations to $2^2\times\sum_{z=0}^6\left(N \atop z\right)$ elements of the many-body evolution operator $U(t)$ resulting from billions of walks on the graph. For $N\gtrsim 100$ we limit our calculations to three excitations and randomly chosen subsets of configurations with four or more excitations.  We typically obtain the many-body wave function $|\psi(t)\rangle$ for 2D lattices with $N \approx 6600$ with moderate computational effort \footnote{Calculations require 18Gbytes of memory but only take up to 5 hours on 4 CPU cores at 2.2GHz.} for arbitrary sets of parameters in the Hamiltonian. We then calculate the order-parameter $\langle T_j\rangle$ and various correlation functions and check their convergence by varying the randomly chosen samples of configurations with $4 \leq \ell \leq 6$. Contributions with $\ell > 6$ are not considered.
\begin{center}
\begin{figure*}[t!]
\includegraphics[width=1\textwidth]{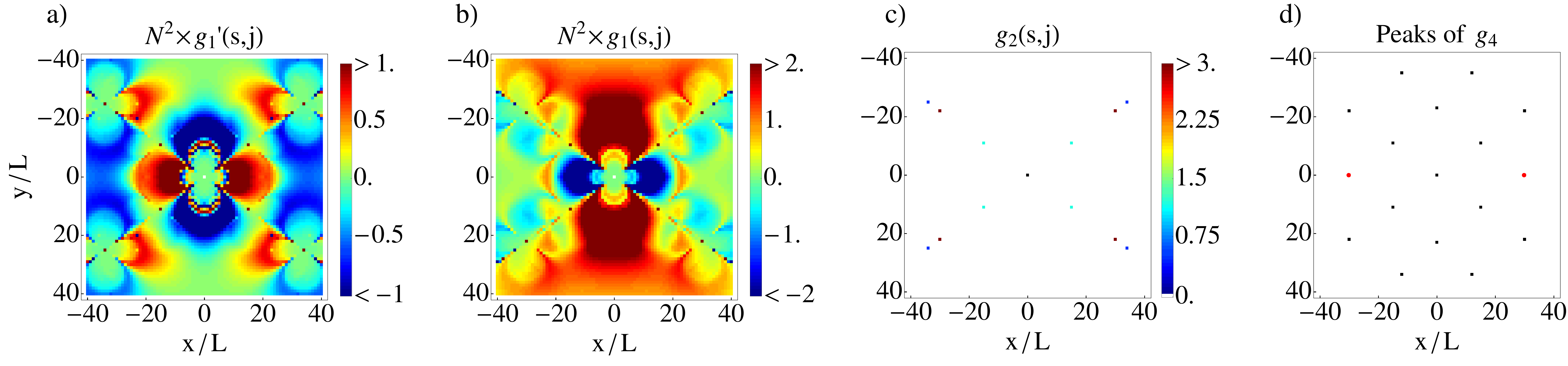}
\caption{(Color online) Correlation-functions multiplied by the number of pairs \textbf{a)} $g_1'(\mathbf{s},\mathbf{j})=\langle T_{\mathbf{s}}^\dagger T^\dagger_{\mathbf{j}}+T_{\mathbf{s}} T_{\mathbf{j}}\rangle-\langle T_{\mathbf{s}}+T_{\mathbf{s}}^\dagger\rangle\langle T_{\mathbf{j}}+T_{\mathbf{j}}^\dagger\rangle$, \textbf{b)} $g_1(\mathbf{s},\mathbf{j})=\langle T_{\mathbf{s}} T^\dagger_{\mathbf{j}}+T_{\mathbf{s}}^\dagger T_{\mathbf{j}}\rangle-\langle T_{\mathbf{s}}+T_{\mathbf{s}}^\dagger\rangle\langle T_{\mathbf{j}}+T_{\mathbf{j}}^\dagger\rangle$ and \textbf{c}) $g_2(\mathbf{s},\mathbf{j})$ over a $N \approx 6600$ atoms square lattice with lattice spacing $L=1.5\mu$m, one atom/pixel, $\mathbf{s}$ is the central one. $g_2(\mathbf{s},\mathbf{j})\simeq0$ except at the few sites where $\mathbf{s}$ and $\mathbf{j}$ form a free pair. \textbf{d}) Locations of the dominant peaks of $g_4=\langle r_{\mathbf{s}} r_{\mathbf{j}} r_{\mathbf{k}} r_{\mathbf{l}}\rangle/\langle r_{\mathbf{s}}\rangle\langle r_{\mathbf{j}}\rangle\langle r_{\mathbf{k}}\rangle\langle r_{\mathbf{l}}\rangle$ (black squares) and $g'_4=\langle g_{\mathbf{s}} r_{\mathbf{j}} r_{\mathbf{k}} r_{\mathbf{l}}\rangle/\langle g_{\mathbf{s}}\rangle\langle r_{\mathbf{j}}\rangle\langle r_{\mathbf{k}}\rangle\langle r_{\mathbf{l}}\rangle$ (red disks) for all $\mathbf{j},\thinspace \mathbf{k}$ forming free pairs with $\mathbf{s}$.
Parameters : Rydberg-state principal number $n= 40$, $\Omega=30$MHz, $\Delta=0$, $\Omega t=8\pi$, $\theta\sim0.43\pi$ and $\phi=\pi/2$, convergence indicates a precision $\sim10^{-4}$.}
\vspace{-5mm}
\label{fig:freepairs}
\end{figure*}
\end{center}

A simple picture of the density-density correlation function $g_2(s,j)=\langle r_sr_j\rangle/\langle r_s\rangle\langle r_j\rangle$ is available at the lowest order, where we find it to be given by
\begin{equation}
\label{eq:g2}
\hspace{-1mm}g_2(s,j)\hspace{-.7mm}=\hspace{-1.1mm}\left(\hspace{-1.2mm}1-\hspace{-.7mm}\frac{\Omega ^2}{\chi_{sj}^2}\hspace{-.2mm} \sin\hspace{-.9mm}\left[\frac{\chi _{sj}}{2}t\right]^2\hspace{-2mm}+\hspace{-.7mm}\frac{\Omega ^2}{\chi ^2}\hspace{-.2mm}\sin\hspace{-.9mm}\left[\frac{ \chi }{2}t\right]^2\hspace{-.7mm}\right)\hspace{-1mm}\frac{\chi^2\hspace{-.2mm}\sin\hspace{-.7mm}\left[\frac{\chi _{sj}}{2}t\right]^2}{\chi_{sj}^{2}\hspace{-.2mm}\sin\hspace{-.7mm}\left[\frac{ \chi }{2}t\right]^2},
\end{equation}
with $\chi^2=\Omega^2+\Delta^2$ and $\chi_{sj}^2=\Omega^2+(\Delta+A_{sj})^2$. This can be understood from conditional-probabilities $g_2(s,j)=(1-\langle r_s/r_j\rangle+\langle r_s/g_j\rangle)\langle r_s/r_j\rangle/\langle r_s/ g_j\rangle$ with e.g.~$\langle r_s/g_j\rangle$ being the probability of finding atom $S$ excited provided $j$ is in the ground state. At this low order no atom is excited in the lattice except for $S$ and $j$ and the conditional-probabilities are expected to be identical to those of the two-atom problem, just as found here. Thus as long as the probability of having more than one excitation in the vicinity of $S$ can be neglected, higher orders should be irrelevant. This result is accurate for up to $\sim2\pi$-pulses in strongly blockaded situations and provides similar values as the $\Omega$-expansion for 1D systems \cite{Stanojevic2008}. Because Eq.(\ref{eq:g2}) is just a two-body picture however, it completely fails to account for correlations between atoms that do not interact directly and where thus four body processes become dominant.

This situation is e.g.~realized in square lattices when the laser angles with respect to the lattice plane $(\theta,\phi)$ fulfil
\begin{equation}
\label{eq:condsol}
3\sin^2(\theta)(j_x\cos(\phi)+ j_y\sin(\phi))^2=j_x^2+j_y^2,
\end{equation}
where $\mathbf{ j}=(j_x,j_y)$ are the lattice coordinates of atom $\mathbf{ j}$ and $\mathbf{s}=(0,0)$. Then the interaction term $A_{\mathbf{sj}}=0$ and in general $|A_{\mathbf{sj'}}|<\Omega$ on the line joining $\mathbf{s}$ and $\mathbf{ j}$. From now on pairs $\mathbf{s},\thinspace \mathbf{j}$ fulfilling Eq.(\ref{eq:condsol}) will be called free pairs. In the presence of such pairs, we observe states with non-zero order-parameter $\langle T_{\mathbf{j}}\rangle$. As shown in Fig. 2 they also display substantial and correlated $g_2(\mathbf{s},\mathbf{j})$ and $g_1'(\mathbf{s},\mathbf{j})$, anticorrelated with $g_1(\mathbf{s},\mathbf{j})$ and all peaking precisely where $\mathbf{s}$ and $\mathbf{j}$ form a free pair. Additionally, we find $\langle r_{\mathbf{s}}\rangle = \sum_{\mathbf{j}}\langle r_{\mathbf{s}}r_{\mathbf{j}}\rangle$ within our numerical accuracy indicating that excitations occur at least in pairs, and this sum being dominated by a $99\%$ contribution from free pairs. By tuning $\theta$ and $\phi$ one can choose which pairs of atoms are free and hence the structure of the correlations.

To further determine the nature of the observed states we compute their pressure $p_{2\textrm{D}} = -(\partial E/\partial\mathcal{A})_{T,Nf_R}$ where $E$ and $\mathcal{A}=NL^2$ are the energy and area of the system, respectively. The derivative is taken at constant temperature $T=0$ and number of excitations $Nf_R$, with $f_R=\sum_{\mathbf{k}}\langle\psi(t_0)| P_{\mathbf{k}}|\psi(t_0)\rangle/N$ the Rydberg fraction. 
To obtain $p_{2D}$ we turn off the laser at time $t_0$ and consider a contraction of the system $L(t)$. The resulting Hamiltonian $H'$ commutes with itself at any time as well as with any of the $P_iP_j$ thus leaving $f_R$ unchanged while $E(t)=\sum_{\mathbf{i},\mathbf{j}}A_{\mathbf{ij}}(t)\langle\psi(t_0)| P_{\mathbf{i}}P_{\mathbf{j}}|\psi(t_0)\rangle$ changes only through $L(t)$. This yields $p_{2\textrm{D}}=(3/2)E/\mathcal{A}$.
As discussed above the sum in the energy is dominated by free pairs for which $A_{\mathbf{ij}}=0$ and we thus find these states to have a very small pressure.
Physically this is because Eq.(\ref{eq:condsol}) is independent of the lattice spacing $L$. Consequently, during a contraction, atoms forming free pairs never interact directly. We expect this small pressure to increase the lifetime of the observed states $\tau\propto p_{2\textrm{D}}^{-1/2}$ compared to states containing interacting excitations.

The presence of both diagonal and off-diagonal long-range order together with a nearly vanishing quantum pressure qualifies the observed states as supersolids \cite{DiagOrder}. Their translational symmetry can only become apparent from $n\geq4$-body correlation-functions as shown in Fig.(\ref{fig:freepairs}-d).  Finally we note that around $\phi=0$ and $\pi/2$, Eq.(\ref{eq:condsol}) is a robust condition as errors of $\Delta\phi$ and $\Delta\theta$ on the laser angles produce a small residual interaction $|A_{\mathbf{ij}}|\lesssim2\sqrt{2}\thinspace\mu_{\mathbf{i}}\mu_{\mathbf{j}}(4\pi\epsilon_0R_{\mathbf{ij}}^{3})^{-1}\Delta\theta$ in free pairs.

In this article we have analysed many-body Rydberg-excited quantum states created by short laser pulses from an atomic Mott insulator of ultracold atoms. We have calculated their pressure, and first, second and fourth order correlation functions to show that these states fulfil the properties of a supersolid for specific laser setups. Our calculations were based on WS which provide a versatile tool for investigating the coherent dynamics of many body quantum systems whose initial wave function is known. WS are part of a generic method to evaluate general matrix functions and can e.g. be extended to analyse continuous-time isotropic quantum random walks on any lattice geometry and be shown to reduce to path-integrals for systems with continuous-only degrees of freedom. 

%
%

\begin{acknowledgments}
This work is supported by a Scatcherd European scholarship and an EPSRC scholarship.\vspace{-3mm}
\end{acknowledgments}

\end{document}